\documentclass[aps,prd,onecolumn,showpacs,groupedaddress, nofootinbib]{revtex4}  
\usepackage{graphicx}
\usepackage{epstopdf}
\usepackage{amsmath}
\usepackage{amsfonts}
\usepackage{amssymb}
\usepackage{appendix}
\usepackage{comment}
\usepackage{bbold}
\usepackage{color}
\usepackage{slashed}
\usepackage{subfigure}
\usepackage{setspace}
\usepackage{footnote}
\usepackage{multirow}

\begin{document}

\singlespacing

{\hfill NUHEP-TH/14-10}

\title{CP-Invariance Violation at Short-Baseline Experiments in $3+1$ Neutrino Scenarios}

\author{Andr\'{e} de Gouv\^{e}a} 
\author{Kevin J. Kelly}
\author{Andrew Kobach}
\affiliation{Northwestern University, Department of Physics \& Astronomy, 2145 Sheridan Road, Evanston, IL 60208, USA}

\begin{abstract}

New neutrino degrees of freedom allow for more sources of CP-invariance violation (CPV). We explore the requirements for accessing CP-odd mixing parameters in the so-called $3+1$ scenario, where one assumes the existence of one extra, mostly sterile neutrino degree of freedom, heavier than the other three mass eigenstates. As a first step, we concentrate on the $\nu_{e}\to\nu_{\mu}$ appearance channel in a hypothetical, upgraded version of the $\nu$STORM proposal. We establish that the optimal baseline for CPV studies depends strongly on the value of $\Delta m^2_{14}$ -- the new mass-squared difference -- and that the ability to observe CPV depends significantly on whether the experiment is performed at the optimal baseline. Even at the optimal baseline, it is very challenging to see CPV in $3+1$ scenarios if one considers only one appearance channel. Full exploration of CPV in short-baseline experiments will require precision measurements of tau-appearance, a challenge significantly beyond what is currently being explored by the experimental neutrino community.

\end{abstract}


\maketitle

\setcounter{equation}{0}
\section{Introduction}
\label{introduction}

The existence of Standard Model (SM) gauge singlet fermions -- sterile neutrinos -- is a very simple and attractive extension to our understanding of fundamental particle physics. Sterile fermions may play a central role when it comes to addressing several of the current outstanding questions, including the dark matter puzzle and the origin of nonzero neutrino masses \cite{Abazajian:2012ys}. 

At the renormalizable level, the only allowed interactions of sterile neutrinos with SM degrees of freedom are those described by Yukawa operators containing left-handed fermions, the Higgs doublet, and the sterile neutrinos. Phenomenologically, this implies that observable sterile-neutrino effects are mostly mediated by the mixing between the active neutrinos ($\nu_e, \nu_{\mu}, \nu_{\tau}$) and the sterile neutrino states. If the sterile neutrino masses are very low (say, below 10~eV), their properties can, almost exclusively,\footnote{Other options include neutrinoless double-beta decay and precision measurements of $\beta$-decay energy spectra. See, for example, \cite{deGouvea:2006gz,Formaggio:2011jg,Rodejohann:2011mu,Esmaili:2012vg}.} be explored via neutrino oscillation experiments.  

Over the last couple of decades, distinct experiments have revealed anomalies that are not consistent with the SM augmented by massive active neutrinos \cite{Aguilar:2001ty,AguilarArevalo:2008rc,Aguilar-Arevalo:2013pmq,Mention:2011rk,Mueller:2011nm,Frekers:2011zz}. These can be interpreted as evidence for more than three neutrinos, with a new oscillation length proportional to a new mass-squared difference around 1~eV$^2$ (for recent analyses, see \cite{Giunti:2012tn,Conrad:2012qt,Kopp:2013vaa,Giunti:2013aea}). Given that the number of active neutrinos is known to be three \cite{Agashe:2014kda}, the extra degrees of freedom must be sterile neutrinos. While our understanding of these so-called short-baseline anomalies remains clouded, there are several experimental proposals aimed at definitively testing the sterile-neutrino interpretation \cite{deGouvea:2013onf}. It is possible that, in five to ten years, experiments will reveal, beyond reasonable doubt, the existence of new neutrino degrees of freedom. Such a monumental discovery would qualitatively impact our understanding of fundamental physics and would invite a new generation of short-baseline neutrino oscillation experiments capable of exploring the new-physics sector. 

Among the properties of the newly-discovered neutrino states would be their couplings to the SM particles, including the probabilities that these would act as $\nu_e,~ \nu_{\mu},~ \nu_{\tau}$, and the relative phases among the new elements of the augmented leptonic mixing matrix. Even if there were only one new neutrino state, new sources of CP-invariance violation (CPV) would become accessible. Given our current understanding of CPV and the potential importance of this phenomenon to some of the basic contemporary particle physics questions, it would be imperative to understand whether, and under what circumstances, these new CPV phenomena are experimentally accessible. 

Here, we discuss the challenges associated with studying CPV in the new-physics sector, assuming that next-generation short-baseline experiments confirm the existence of new neutrino states with parameters that are approximately consistent with those indicated by the sterile-neutrino interpretation to the current short-baseline anomalies. We restrict our discussion to the case of only one accessible new neutrino state. According to \cite{Kopp:2013vaa}, the assumption that there are at least two accessible states might be a better fit to the short-baseline data. However, if there are two (or more) sterile neutrinos, CPV present in the interference between the two (or more) new oscillation frequencies may already have manifested itself in the current generation of short-baseline experiments \cite{Karagiorgi:2006jf,Giunti:2012tn,Conrad:2012qt,Kopp:2013vaa,Giunti:2013aea}, and is hence more-or-less straight forward to observe. On the other hand, if only one new neutrino state is accessible, CPV will manifest itself in the interference between the new mass-squared difference and the known atmospheric and solar ones, a phenomenon which depends only on a few new-physics parameters and may turn out to be much more challenging to explore experimentally.  These interference effects are very small and virtually impossible to observe in current and proposed experimental setups, which can safely neglect them. We discuss this further in Sections~\ref{parameterization} and~\ref{sec:exp}. 

In Sec.~\ref{parameterization}, we discuss $3+1$ oscillations, concentrating on experimental circumstances where only two of the three independent oscillation frequencies are accessible. We present the relevant oscillation probabilities and discuss which parameters can be measured and what are the different sources of CPV. In Sec.~\ref{sec:exp}, we discuss the requirements for observing $3+1$ CPV in short baseline experiments, and explore the capabilities of a concrete next-next-generation experimental setup, inspired by the $\nu$STORM proposal \cite{Adey:2014rfv}, to study CPV in a high-statistics, high-resolution short-baseline experiment. In Sec.~\ref{sec:conclusion}, we summarize our results and offer some concluding thoughts. 

\setcounter{equation}{0}
\section{Neutrino Oscillations at Short Baselines}
\label{parameterization}

Under the hypothesis that there are four neutrino states $\nu_{i}$, $i=1,2,3,4$, $P_{\alpha\beta}(E_{\nu},L)$ -- the probability that a $\nu_\alpha$ flavor eigenstate with energy $E_{\nu}$ is detected as a $\nu_\beta$ flavor eigenstate, $\alpha,\beta=e,\mu,\tau$, after it propagates a distance $L$ -- is given by the absolute value squared of the oscillation amplitude $\mathcal{A}_{\alpha\beta}$. For $\alpha \neq \beta$, 
\begin{equation}
\label{Amplitude}
\mathcal{A}_{\alpha\beta} = U_{\alpha 2}U_{\beta 2}^* \left(e^{-i\Delta_{12}}-1\right) + U_{\alpha 3}U_{\beta3 }^* \left(e^{-i\Delta_{13}}-1\right) + U_{\alpha 4} U_{\beta 4}^* \left(e^{-i\Delta_{14}}-1\right).
\end{equation}
Here, $\Delta_{ij} \equiv 2.54(\Delta m_{ij}^2/1 \text{ eV}^2)(L/\text{km})(\text{GeV}/E_\nu)$ and $\Delta m_{ij}^2 \equiv m_j^2 - m_i^2$, where $m_i$ are the neutrino masses, $i,j=1,2,3,4$. $U_{\alpha i}$ are the elements of the unitary $4\times 4$ neutrino mixing matrix, $\alpha=e,\mu,\tau,s$, where $s$ stands for the sterile neutrino. Note that the $U_{si}$ elements are not accessible to experiments, assuming there are no interactions directly sensitive to the sterile neutrino state.

We presume that the matrix elements $U_{\alpha i}$ and the values of $\Delta m_{ij}^2$ are such that they fit the existing long-baseline neutrino data for $i=1,2,3$, $\alpha=e,\mu,\tau$ \cite{Capozzi:2013csa,Forero:2014bxa,Gonzalez-Garcia:2014bfa}. We further assume that next-generation short-baseline neutrino oscillation experiments will confirm the existence of one new mass-squared difference, $|\Delta m_{14}|^2\sim |\Delta m_{24}^2|\sim|\Delta m_{34}^2|\gg |\Delta m^2_{13}|,$ $\Delta m^2_{12}$, consistent with the sterile neutrino interpretation of  the short-baseline anomalies~\cite{Giunti:2012tn,Conrad:2012qt,Kopp:2013vaa,Giunti:2013aea}. Hence, we assume 
\begin{equation}
\label{MassRange}
\Delta m_{14}^2 \in [0.1, 10]\text{ eV}^2,
\end{equation}
and will only consider the mass ordering where $m_4^2\gg m_3^2,m_2^2,m_1^2$. The effective mixing angle $|U_{e4}||U_{\mu 4}|$ is assumed to lie within the range
\begin{equation}
\label{MixAngleRange}
|U_{e4}||U_{\mu_4}| \in [0.01, 0.15].
\end{equation}
Note that this assumption is consistent with $|U_{e4}||U_{\mu_4}|\sim |U_{e3}|^2\simeq 0.02$. 

We parameterize the elements of the $4\times 4$ unitary transformation $U$ as  (ignoring potentially physical, but irrelevant-for-oscillations, Majorana phases)
\begin{eqnarray}
\label{Ue2}U_{e2} &=& s_{12}c_{13}c_{14}, \\
U_{e3} &=& e^{-i\delta}c_{14}s_{13}, \\
U_{e4} &=& s_{14}e^{-i\delta_1}, \\
U_{\mu 2} &=& c_{24}\left(c_{12}c_{23}-e^{i\delta}s_{12}s_{13}s_{23}\right) - e^{i(\delta_1-\delta_2)}c_{13}s_{12}s_{14}s_{24}, \\
\label{Umu3}U_{\mu 3} &=& c_{13}c_{24}s_{23}-e^{i(\delta_1 -\delta_2 - \delta)}s_{13}s_{14}s_{24},\\
U_{\mu 4} &=& s_{24}c_{14}e^{-i\delta_2},\\
U_{\tau 2} &=& c_{34}\left(-e^{i\delta}c_{23}s_{12}s_{13}-c_{12}s_{23}\right)-e^{i\delta_1}c_{13}c_{24}s_{12}s_{14}s_{34}\nonumber \\
&&-e^{i\delta_2}\left(c_{12}c_{23}-e^{i\delta}s_{12}s_{13}s_{23}\right)s_{24}s_{34,}\\
\label{Utau3}U_{\tau 3} &=& c_{13}c_{23}c_{34}-e^{i(\delta_1-\delta)}c_{24}s_{13}s_{14}s_{34}-e^{i\delta_2}c_{13}s_{23}s_{24}s_{34},\\
\label{Utau4}U_{\tau 4} &=& s_{34}c_{14}c_{24},
\end{eqnarray}
where $s_{ij} \equiv \sin{\theta_{ij}},$ $ c_{ij} \equiv \cos{\theta_{ij}}$, $(i,j = 1, 2, 3, 4)$. The matrix elements depend on six mixing angles ($\theta_{12},\theta_{13},\theta_{23},\theta_{14},\theta_{24},\theta_{34}$) and three CP-odd phases ($\delta,\delta_1,\delta_2$). The elements not listed here can be determined by imposing unitarity conditions on $U$.  

If one of the new mixing angles -- $\theta_{14},$ $\theta_{24},$ $\theta_{34}$ -- were to vanish, one of the new CP-odd phases -- $\delta_1$ and $\delta_2$, or combinations thereof -- would become non-physical, as expected. A similar phenomenon would be observed if any of the mass-squared differences were to vanish. While we know (or assume) that all $\Delta m^2_{ij}\neq 0$, their effects might still be unobservable. Since the mass-squared differences are quite hierarchical -- $\Delta m^2_{12}\ll |\Delta m^2_{13}|\ll \Delta m^2_{14}$ -- we examine this issue in more detail.

We will consider experiments that probe $P_{\alpha\beta}$ when $E_\nu= \mathcal{O}(1\text{ GeV})$ and baselines $L=\mathcal{O}(1\text{ km})$, i.e., $L/E_\nu \sim 1$~km/GeV, such that $\Delta_{14}\sim 1$.\footnote{The other possibility is to aim at the atmospheric oscillation, $L/E_\nu \sim 1000$ km/GeV, such that $|\Delta_{13}|\sim 1$. We do not consider this case here. When $|\Delta_{13}|\sim 1$, the fast oscillations associated with the (mostly) sterile neutrino tend to average out, rendering the study of CPV very challenging, because the effects of CPV are most apparent when comparing different values of $L/E$. If the new oscillations do ``average out,'' then this does not necessarily remove the possibility of indirectly exploring CPV phenomena; the combination of results from multiple experiments can be used to measure CPV effects in a $3+1$ scenario \citep{Hollander:2014iha, Klop:2014ima}.}  Under these circumstances,  $\Delta_{12}=\mathcal{O}(10^{-5})$ and $|\Delta_{13}|=\mathcal{O}(10^{-3})$. 
With the above information in mind, we revisit Eq.~(\ref{Amplitude}), taking into account that $(e^{-i\Delta_{12,13}} - 1) \simeq -i\Delta_{12,13}$.  To illustrate the relative size of terms in Eq.~(\ref{Amplitude}), we define $(\mathcal{R}_{\alpha\beta})_{ij}$ as the ratio of the ``$1i$'' to the ``$1j$'' contribution to $\mathcal{A}_{\alpha\beta},$\footnote{The approximation $\Delta_{14}\sim1$ implies that the ``14'' term is of order $|U_{\alpha 4}U_{\beta4}^*|$ and $L/E_{\nu}\sim(\Delta m^2_{14})^{-1}$. }
\begin{equation}
\label{Ratio}
\left(\mathcal{R}_{\alpha\beta}\right)_{ij} \sim \frac{|U_{\alpha i}U^*_{\beta i}|\Delta m^2_{1i}}{|U_{\alpha j}U^*_{\beta j}|\Delta m^2_{1j}},
\end{equation}
for $i,j=2,3,4$. For all $\alpha$ and $\beta$, $\left(\mathcal{R}_{\alpha\beta}\right)_{23}$ and $\left(\mathcal{R}_{\alpha\beta}\right)_{24}$ are small. For example, even though $|U_{e3}| = \sin{\theta_{13}} \simeq 0.15$ is small compared to $|U_{e2}| \simeq 0.55$, the ratio between $\Delta m^2_{12}$ and $\Delta m^2_{13}$ is such that $\left(\mathcal{R}_{e\mu}\right)_{23} \simeq 0.1$. Additionally, considering a new mass splitting in agreement with Eq.~(\ref{MassRange}) and mixing angles in agreement with Eq.~(\ref{MixAngleRange}), the ratio $\left(\mathcal{R}_{e\mu}\right)_{24} \in [10^{-5}, 10^{-2}]$. Thus, it is practical to set $\Delta_{12} = 0$, which is an approximation we make henceforth. Furthermore, since we only consider $|U_{e4}U^*_{\mu_4}|\Delta m^2_{14}\gtrsim 8\times 10^{-3}$~eV$^2$, which is approximately four times larger than $|\Delta m^2_{13}|$, $\left(\mathcal{R}_{e\mu}\right)_{34}\lesssim 10^{-1}$ is also small. In summary, if the oscillation interpretation of the short-baseline anomalies is correct, in experiments performed at $L$ and $E_{\nu}$ values where $\Delta_{14}\sim 1$, solar contributions are irrelevant and atmospheric contributions are small, at least in the $e\mu$ sector.  If $|U_{\tau 4}|\sim|U_{\mu 4}|,$ $|U_{e4}|$,\footnote{Currently, there is very little experimental information regarding the $\tau$ sector.} the same is approximately true of the $e\tau$ and $\mu\tau$ sectors, even when one takes into account that $|U_{\tau 3}|$ is several times larger than $|U_{e3}|$.

\setcounter{footnote}{0}

In the limit $\Delta m^2_{12}\to 0$, we ``lose'' the angle $\theta_{12}$ and the CP-odd phase $\delta$,\footnote{When $\Delta_{12}=0$ in Eq.~(\ref{Amplitude}), one is not sensitive to any $U_{\alpha 2}$ and hence the value of $\theta_{12}$. This, in turn, implies that the amplitude is consistent with any value of $\theta_{12}$, including $\theta_{12}=0$. When $\theta_{12}=0$, one of the CP-odd phases is unphysical.}  and the oscillation probabilities depend on five angles ($\theta_{13}, \theta_{23}, \theta_{14}, \theta_{24}, \theta_{34}$) and two independent CP-odd phases, which we define as $\psi_s \equiv \delta_1 - \delta$ and $\phi_s \equiv (\delta_1 - \delta) - \delta_2$. Taking, in addition, the limit $\Delta m^2_{13}\to 0$, the oscillation probabilities depend on three angles ($\theta_{14},\theta_{24},\theta_{34}$) and zero physical CP-odd phases.\footnote{In general, in the limit where $j$ mass-squared splittings vanish, some of the observables that parameterize $U$ become unphysical. The total number of angles $N_\theta$ and phases $N_\delta$ that determine $P_{\alpha\beta}$ in this case are
\begin{eqnarray}
\label{NAngles}N_\theta &=& n(n-1)/2 -j(j+1)/2,\\
\label{NPhases}N_\delta &=& (n-1)(n-2)/2 - j(j+1)/2,
\end{eqnarray}
assuming there are $n$ neutrino states.} The latter limit is the one usually considered in the analyses of short-baseline experiments~\citep{Adey:2014rfv,Giunti:2012tn,Conrad:2012qt,Kopp:2013vaa,Giunti:2013aea}. 

When assuming there is only one relevant sterile neutrino, therefore, the study of CPV at short-baseline experiments requires sensitivity to the small $\Delta_{13}$ effects. Since disappearance channels are CP-invariant as a consequence of the CPT-theorem, we concentrate on the appearance channels.\footnote{One could try to infer that CP-invariance is violated by comparing different disappearance channels and fitting them to the $3+1$ oscillation hypothesis. We do not explore this possibility here.} Taking advantage of what is known (or assumed) about the mixing parameters, we can further simplify the oscillation expressions. In detail, we approximate $U_{\mu 3} = c_{13}c_{24}s_{23}$ and $U_{\tau 3} = c_{13}c_{23}c_{34}$, since the subleading terms in Eqs.~(\ref{Umu3}) and (\ref{Utau3}) are $\mathcal{O}(10^{-2})$, so the appearance probabilities can be written as
\begin{eqnarray}
\label{probmutau}
P_{\mu\tau} &\simeq& 4c_{14}^4 s_{24}^2 c_{24}^2 s_{34}^2 \sin^2{\left(\frac{\Delta_{14}}{2}\right)} \nonumber \\
&+&8 c_{13}^2 s_{23} c_{23} c_{14}^2 s_{24} c_{24}^2 s_{34} c_{34}\sin{\left(\frac{\Delta_{13}}{2}\right)}\sin{\left(\frac{\Delta_{14}}{2}\right)}\cos{\left(\frac{\Delta_{14}}{2}+\psi_s-\phi_s\right)} \nonumber \\
&+&4c_{13}^4 s_{23}^2 c_{23}^2 c_{24}^2 c_{34}^2 \sin^2{\left(\frac{\Delta_{13}}{2}\right)},
\\
\label{probetau}
P_{e\tau} &\simeq& 4s_{14}^2 c_{14}^2 c_{24}^2 s_{34}^2 \sin^2{\left(\frac{\Delta_{14}}{2}\right)} \nonumber \\
&+&8 s_{13}c_{13}c_{23} s_{14} c_{14}^2 c_{24} s_{34} \sin{\left(\frac{\Delta_{13}}{2}\right)}\sin{\left(\frac{\Delta_{14}}{2}\right)}\cos{\left(\frac{\Delta_{14}}{2}+\psi_s\right)} \nonumber \\
&+&4 s_{13}^2 c_{13}^2 c_{23}^2 c_{14}^2 c_{34}^2\sin^2{\left(\frac{\Delta_{13}}{2}\right)},
\\
\label{probemu}
P_{e\mu} &\simeq& 4s_{14}^2 c_{14}^2 s_{24}^2 \sin^2{\left(\frac{\Delta_{14}}{2}\right)}\nonumber \\
&+& 8 s_{13}^2 c_{13} s_{23} s_{14} c_{14}^2 s_{24} c_{24} \sin{\left(\frac{\Delta_{13}}{2}\right)} \sin{\left(\frac{\Delta_{14}}{2}\right)} \cos{\left(\frac{\Delta_{14}}{2} + \phi_s\right)}\nonumber \\
&+& 4 s_{13}^2 c_{13}^2 s_{23}^2 c_{14}^2 c_{24}^2 \sin^2{\left(\frac{\Delta_{13}}{2}\right)}.
\end{eqnarray}
The CP-conjugate and T-conjugate channels are obtained by changing the sign of the CP-odd phases $\psi_s$ and $\phi_s$, i.e., $P_{\bar{\alpha}\bar{\beta}}(\phi_s,\psi_s)=P_{\alpha\beta}(-\phi_s,-\psi_s)$ and $P_{\beta\alpha}(\phi_s,\psi_s)=P_{\alpha\beta}(-\phi_s,-\psi_s)$.  The explicit CPV effects that render $P_{\alpha\beta}\neq P_{\bar{\alpha}\bar{\beta}}$ are contained in the interference between the $\Delta_{13}$ and the $\Delta_{14}$ terms in Eq.~(\ref{Amplitude}). In each $P_{\alpha\beta}$, the ``14-squared'' term is dominant, the interference term is the next-to-leading term, followed by the ``13-squared'' term, which is smallest. The measurement of two different appearance channels is required in order to determine the two independent CP-odd phases and, in principle, the measurement of the third appearance channel would serve as a nontrivial test of the $3+1$ hypothesis. 

We assume that experiments will reveal that neither $\theta_{14}$ nor $\theta_{24}$ is very small, but anticipate learning very little about $\theta_{34}$, which is linked to $U_{\tau 4}$ and tau-appearance. Furthermore, working with taus is extremely challenging. It requires ``detection'' center-of-mass energies larger than the tau mass, and detectors capable of identifying taus with nonzero efficiency. Henceforth, we utilize exclusively the appearance oscillation probability $P_{e\mu}$, returning to taus in the concluding statements. In the range of values for $\theta_{14}$ and $\theta_{24}$ satisfying Eq.~(\ref{MixAngleRange}), the oscillation probability in Eq.~(\ref{probemu}) is approximately degenerate under interchange of $\theta_{14} \leftrightarrow \theta_{24}$; thus, we choose to simplify the parameterization by taking $\theta_s \equiv \theta_{14} = \theta_{24}$, and rewrite Eq.~(\ref{probemu}) as
\begin{eqnarray}
\label{probemus}
P_{e\mu} &\simeq& 4s_{s}^4 c_{s}^2 \sin^2{\left(\frac{\Delta_{14}}{2}\right)}\nonumber \\
&+& 8 s_{13}^2 c_{13} s_{23} s_{s}^2 c_{s}^3 \sin{\left(\frac{\Delta_{13}}{2}\right)} \sin{\left(\frac{\Delta_{14}}{2}\right)} \cos{\left(\frac{\Delta_{14}}{2} + \phi_s\right)}\nonumber \\
&+& 4 s_{13}^2 c_{13}^2 s_{23}^2 c_{s}^4 \sin^2{\left(\frac{\Delta_{13}}{2}\right)},
\end{eqnarray}
where $c_s \equiv \cos{\theta_s}$ and $s_s \equiv  \sin{\theta_s}$. Our results are not sensitive to this assumption. Instead, the combined analyses of $\nu_e$ or $\nu_{\mu}$ appearance and $\nu_{\mu}$ or $\nu_e$ disappearance can distinguish $\theta_{14}$ effects from those of $\theta_{24}$. We do not pursue such an analysis here. Finally, Eq.~(\ref{probemus}) (and Eq.~(\ref{probemu})) is invariant under $\Delta m_{13}^2 \to -\Delta m^2_{13}$ and  $\phi_s \to \phi_s + \pi$. For this reason, we assume henceforth that the sign of $\Delta m^2_{13}$ is positive. Our results are still valid if the mass-hierarchy turns out to be inverted, but for the shifted value of $\phi_s$.

\setcounter{equation}{0}
\section{Experimental Sensitivity to CP-Violating Phases}
\label{sec:exp}

We investigate the capability of next-next-generation experimental efforts to see $3+1$ CPV by simulating short-baseline experiments based on the $\nu$STORM proposal~\cite{Adey:2014rfv}. According to the discussion in Section~\ref{parameterization}, $3+1$ CPV effects are quite small and therefore require large statistics, excellent control of systematics, and very good energy resolution. All of these are potentially within reach of future neutrino experiments with beams from muon decay in flight.  Other ideas for future experiments should be explored, including pion-decay-at-rest ``beams,'' similar to, for example, DAE$\delta$ALUS \cite{Alonso:2010fs}.  

The $\nu$STORM proposal is designed to measure the values of $|U_{e4}|^2|U_{\mu4}|^2$ and $\Delta m_{14}^2$ in the $\nu_e \rightarrow \nu_\mu$ appearance channel for a $3+1$ scenario~\cite{Adey:2014rfv}.  Unlike pion-decay-in-flight long-baseline experiments that investigate the process $\nu_\mu \rightarrow \nu_e$,  $\nu$STORM  uses $\nu_e$ and $\overline{\nu}_\mu$ from the decay of stored $\mu^+$ to produce two neutrino beams.  A detector with a strong magnetic field allows for $\mathcal{O}(1\%)$ energy resolution and powerful discrimination between detecting $\mu^+$ from $\overline{\nu}_\mu\rightarrow \overline{\nu}_\mu$ and $\mu^-$ from $\nu_e \rightarrow \nu_\mu$, which dramatically reduces beam-related backgrounds. 

\begin{figure}[htbp]
\begin{center}
\includegraphics[width=0.7\textwidth]{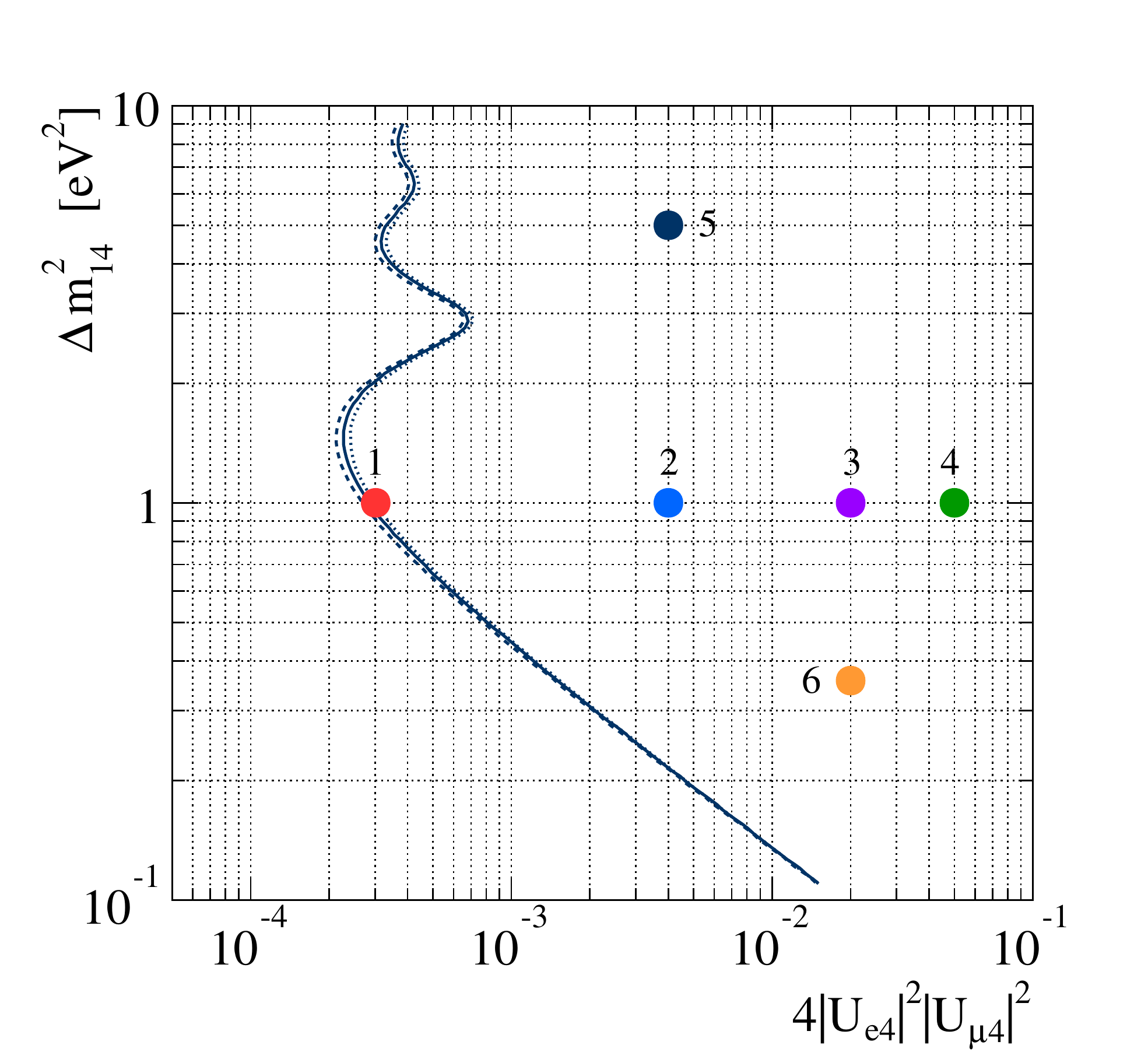}
\caption{The solid line is the 99\% CL sensitivity using the $\nu_e \rightarrow \nu_\mu$ channel at $\nu$STORM, assuming $\Delta m^2_{13}=0$, in agreement with~\cite{Adey:2014rfv}. The dashed and dotted lines correspond to the 99\% CL sensitivity if atmospheric effects are taken into account (see Eq.~(\ref{probemus})), when $\phi_s=-\pi/2$ and $\phi_s=\pi/2$, respectively. The six numbered points correspond to values of $|U_{e4}|^2|U_{\mu4}|^2$ and $\Delta m_{14}^2$ used in this analysis to discuss the measurement of the value of $\phi_s$. }
\label{nustormplot}
\end{center}
\end{figure}

First, we reproduce the $\nu$STORM analysis in~\cite{Adey:2014rfv} using similar flux, cross section, detector design, background rate, signal and background efficiency, and systematic uncertainties (1\% and 10\% associated with signal and background normalizations, respectively)~\cite{Kyberd:2012iz, Formaggio:2013kya}. This analysis is performed in the limit $\Delta m^2_{13}\to0$. Our results, depicted by the solid line in Fig.~\ref{nustormplot}, agree with those from~\cite{Adey:2014rfv} and illustrate that the $\nu$STORM experiment with a 1.3 kt detector at $L=2$~km would be able to constrain $4|U_{e4}|^2|U_{\mu4}|^2 < \mathcal{O}(10^{-4} - 10^{-3})$ at 99\% CL for $\Delta m_{14}^2 \gtrsim 0.5 \text{ eV}^2$ after 10 years of running.    
If instead the $3+1$ scenario were confirmed by the $\nu$STORM experiment, the precision with which $\nu$STORM could measure $|U_{e4}|^2|U_{\mu4}|^2$ and $\Delta m_{14}^2$  strongly depends on their physical values, as shown in Table~\ref{table:nustormprecision}.  

\begin{table}[h]
\centering
\begin{tabular}{| c | c  | c | c |}
\hline
 & \multirow{2}{*}{Physical Value} & $\nu$STORM  & $\nu$STORM+    \\ 
 & & Precision & Precision \\ \hline \hline
\multirow{2}{*}{Point 1} & $4|U_{e4}|^2|U_{\mu4}|^2=3\times10^{-4}$ & $\mathcal{O}(100\%)$ & $\mathcal{O}(15\%)$  \\ 
& $\Delta m_{14}^2 = 1.0 \text{ eV}^2$ & $\mathcal{O}(100\%)$ & $\mathcal{O}(1\%)$ \\ \hline
\multirow{2}{*}{Point 2} & $4|U_{e4}|^2|U_{\mu4}|^2=4\times10^{-3}$ & $\mathcal{O}(25\%-100\%)$ & $\mathcal{O}(5\%)$ \\ 
& $\Delta m_{14}^2 = 1.0 \text{ eV}^2$ & $\mathcal{O}(30\%)$ & $\mathcal{O}(0.5\%)$ \\ \hline
\multirow{2}{*}{Point 3} & $4|U_{e4}|^2|U_{\mu4}|^2=2\times10^{-2}$ & $\mathcal{O}(20\%)$ & $\mathcal{O}(1\%)$ \\ 
& $\Delta m_{14}^2 = 1.0 \text{ eV}^2$ & $\mathcal{O}(15\%)$ & $\mathcal{O}(0.1\%)$\\ \hline
\multirow{2}{*}{Point 4} & $4|U_{e4}|^2|U_{\mu4}|^2=5\times10^{-2}$ & $\mathcal{O}(10\%)$ & $\mathcal{O}(1\%)$ \\ 
& $\Delta m_{14}^2 = 1.0 \text{ eV}^2$ & $\mathcal{O}(10\%)$ & $\mathcal{O}(0.1\%)$ \\ \hline
\multirow{2}{*}{Point 5} & $4|U_{e4}|^2|U_{\mu4}|^2=4\times10^{-3}$ & $\mathcal{O}(20\%)$ & $\mathcal{O}(1\%)$ \\ 
& $\Delta m_{14}^2 = 5.0 \text{ eV}^2$ & $\mathcal{O}(1\%)$ & $\mathcal{O}(0.05\%)$ \\ \hline
\multirow{2}{*}{Point 6} & $4|U_{e4}|^2|U_{\mu4}|^2=2\times10^{-2}$ & $\mathcal{O}(100\%)$ & $\mathcal{O}(5\%)$ \\ 
& $\Delta m_{14}^2 = 0.35 \text{ eV}^2$ & $\mathcal{O}(100\%)$ & $\mathcal{O}(0.5\%)$ \\ \hline
\end{tabular}
\caption{The position of the six colored points in Fig.~\ref{nustormplot} and the approximate 95\% CL expected precisions with which $\nu$STORM and $\nu$STORM+ can measure $4|U_{e4}|^2|U_{\mu4}|^2$ and $\Delta m_{14}^2$.  While the baseline of $\nu$STORM is 2 km, the baseline of $\nu$STORM+ is optimized to measure CPV by requiring $\Delta m_{14}^2 L = 11.5$ eV$^2\cdot$km with 1000 times more statistics, for the same baseline, than $\nu$STORM. }
\label{table:nustormprecision}
\end{table}

To illustrate the effects of CPV at $\nu$STORM, we recalculate the exclusion limits with the full expression in Eq.~(\ref{probemus}), when $\phi_s= -\pi/2$ and $\pi/2$.\footnote{While we use Eq.~(\ref{probemus}), we convert $\theta_s$ into $|U_{e4}||U_{\mu4}|$ for comparison purposes: $\sin^2\theta_s\cos\theta_s\equiv|U_{e4}||U_{\mu4}|$.} These results are depicted by the dashed and dotted lines in Fig.~\ref{nustormplot}, respectively.  Small differences in the limits occur; the difference between them is of the same order, around one percent, as the effects of the systematic uncertainties outlined in~\cite{Adey:2014rfv}.  To measure CPV, more statistics and an optimal choice of the baseline are required.

In order to investigate what is necessary to measure CPV in the $3+1$ scenario, we consider a dramatically upgraded version of the $\nu$STORM proposal, increasing the data sample -- for the same baseline -- by a factor of 1000 with respect to~\cite{Adey:2014rfv}. This could be achieved if, for example, the beam flux were ten times larger ($\sim 10^{22}$ protons on target) over 10 years and the detector mass were 130~kt.  We will refer to this experiment as $\nu$STORM+. While the proposed beam power and detector mass are outside the realm of possibilities today, they are not entirely outlandish. For comparison purposes, the proposed, and recently approved, India-Based Neutrino Observatory is a 51~kt magnetized iron calorimeter \cite{Kaur:2014rfa}. On the other hand, the proton driver for the proposed Neutrino Factory is planned to deliver $10^{22}$ protons on target per $10^7$ seconds \cite{Weng:2006uz}.  $\nu$STORM+ would accumulate a large enough data sample such that the values of $|U_{e4}|^2|U_{\mu4}|^2$ and $\Delta m_{14}^2$ would be measured very precisely, as displayed in Table~\ref{table:nustormprecision}, at which point the value of $\phi_s$ would begin to induce observable changes to the oscillation probability. 

To analyze our simulated data, we make use of the $\chi^2$ function
\begin{equation}
\chi^2\left(\Delta m^2_{14},\theta_s,\phi_s | \Delta m^{2\star}_{14}, \theta_s^{\star}, \phi_s^{\star}\right) = \displaystyle\sum_i^{\text{bins}}  \frac{\left[N^\text{data}_i(\Delta m^{2\star}_{14},\theta_s^{\star},\phi_s^{\star}) - N^\text{hyp}_i(\Delta m^2_{14}, \theta_s, \phi_s)\right]^2}{N^\text{hyp}_i(\Delta m^2_{14}, \theta_s, \phi_s)},
\label{Chi2}
\end{equation}
where $N^\text{data}_i$ and $N^\text{hyp}_i$ are the measured and expected number of events in energy bin $i$, respectively, and $\Delta m^{2\star}_{14},\theta^{\star}_s,\phi^{\star}_s$ and $\Delta m^2_{14},\theta_s,\phi_s$ are physical (i.e.~input) and hypothetical values of the new mass-squared difference, mixing angle, and CP-odd phase, respectively. Strictly speaking, the $\chi^2$ function also depends on $\theta_{13},\theta_{23},$ and $\Delta m_{13}^2$. We assume, however, that $\theta_{13},\theta_{23},$ and $\Delta m_{13}^2$ will be measured with sufficient precision such that the $\chi^2$ function above, once marginalized over $\theta_{13},\theta_{23},$ and $\Delta m_{13}^2$, is sufficiently indistinguishable from its expression for the best-fit values of $\theta_{13},\theta_{23},$ and $\Delta m_{13}^2$, which we take to be the ones in~\cite{Agashe:2014kda}. Indeed, if one includes the current central values and uncertainties on these parameters \cite{Capozzi:2013csa,Forero:2014bxa,Gonzalez-Garcia:2014bfa}, the oscillation probability $P_{e\mu}$ and the $\chi^2$ function change only at the sub-percent level. We can safely presume that current and future experiments will increase the precision with which these parameters are known before $\nu$STORM$+$ exists. We also considered the possibility that the uncertainties are significantly larger -- 5\% and 50\% associated with the normalization of the signal and background, respectively. The effect of inflating the uncertainties, as far as all results presented henceforth, is negligible.

The value of $N_i$ is determined by integrating, over the bin width,
\begin{equation}
\label{dNdE}
\frac{dN}{dE} =  \Delta t \cdot \Phi(E) \cdot \sigma(E) \cdot \epsilon(E) \cdot P_{e\mu}(E).
\end{equation}
Here $\Phi$, $\sigma$, $\epsilon$, $P_{e\mu}$, and $\Delta t$ are the flux, cross section, efficiency, $\nu_e \rightarrow \nu_\mu$ oscillation probability, and the amount of time the experiment runs, respectively. From Eqs.~(\ref{Chi2}) and (\ref{dNdE}), we see that $\chi^2$ depends linearly on $\Delta t,$ $\Phi(E),$ $\sigma(E)$, and $\epsilon(E)$. 

We can understand, semi-quantitatively, the sensitivity to CPV by analyzing Eq.~(\ref{Chi2}) in more detail and making a few simplifying assumptions. We consider that $N_i$ can be approximated by evaluating $dN/dE$ at the value of the energy corresponding to the center of bin $i$, $E_i$. Taking into account that $\Delta_{13} \ll 1$, and, only for the sake of this discussion, fixing $\Delta m^2_{14}$ and $\theta_s$ to their input values, i.e., setting $\Delta m^2_{14}=\Delta m^{2\star}_{14},\theta_s=\theta_s^{\star}$,
\begin{equation}
\label{Chi2Simp}
\chi^2\left(\phi_s | \phi^{\star}_s\right) \propto \Delta t\sum_{i}^\text{bins} L^2\Phi(E_i)\sigma(E_i)\epsilon(E_i)\left(\frac{s_s^2c_s^4}{s_s^2+ALc_{s}\cos{\left(\frac{\Delta_{14}}{2}+\phi_s\right)}/\sin{\left(\frac{\Delta_{14}}{2}\right)}+BL^2c_s^4}\right)f_i(\phi_s,\phi^{\star}_s),
\end{equation}
where
\begin{equation}
f_i(\phi_s,\phi^{\star}_s)=\left[\cos{\left(\frac{\Delta_{14}}{2}+\phi_s\right)}-\cos{\left(\frac{\Delta_{14}}{2}+\phi^{\star}_s\right)}\right]^2,
\end{equation}
$A\sim \mathcal{O}(10^{-4}\text{ km}^{-1})$, and $B\sim \mathcal{O}(10^{-8}\text{ km}^{-2})$. Even though we are interested in $L\lesssim 100$ km, we preserve the term involving $B$ in the event that the $L/E$-dependent coefficient of $A$ is vanishingly small.

The bins with the largest number of events $N_i$, and therefore the most statistical power, will contribute the most to the value of $\chi^2$. These correspond to the peaks of $P_{e\mu}(E)$, approximately where $\sin{\left(\frac{\Delta_{14}}{2}\right)} = \pm1$. Fixing $\sin{\left(\frac{\Delta_{14}}{2}\right)} = \pm 1$, we can further simplify Eq.~(\ref{Chi2Simp}) down to its dominant contributions
\begin{equation}
\label{Chi2Dom}
\chi^2\left(\phi_s | \phi^{\star}_s\right) \propto \Delta t \sum_{s{\left(\frac{\Delta_{14}}{2}\right)} = \pm 1} L^2\Phi(E_i)\sigma(E_i)\epsilon(E_i)\left(\frac{s_s^2c_s^4}{s_s^2+CLc_s+BL^2c_s^4}\right)g(\phi_s,\phi^{\star}_s),
\end{equation}
where the sum is restricted to the bins where the approximation $\sin{\left(\frac{\Delta_{14}}{2}\right)} = \pm1$ is good, $|C| \lesssim \mathcal{O}(10^{-4}\text{ km}^{-1})$ and
\begin{equation}
g(\phi_s,\phi^{\star}_s) = \left(\sin{\phi_s}-\sin{\phi^{\star}_s}\right)^2.
\end{equation}
Eqs.~(\ref{Chi2Simp}) and (\ref{Chi2Dom}) allow one to conclude the following:
\begin{itemize}
\item Because the flux, $\Phi$, scales like $1/L^2$ and $B$ and $C$ are small numbers, the most significant dependence on $L$ comes in the product $\Delta m_{14}^2 L$.  Therefore, the ability to measure $\phi_s$ is, to a good approximation, greatest for some constant value of the product $\Delta m_{14}^2 L$.  
\item The sensitivity for measuring the value of $\phi_s$ is linearly dependent on the power of the beam, size of the detector, and the amount of time that the experiment runs.
\item The value of $\phi_s$ is easiest to measure when $\theta_s \sim 0.18$, i.e., $4|U_{e4}|^2|U_{\mu4}|^2\sim 4\times10^{-3}$.  Around this value, $\chi^2$ falls off slowly for $\theta_s > 0.18$ and falls off rapidly for $\theta_s < 0.10$.  This is apparent by analyzing the term $(s_s^2c_s^4)/(s_s^2+CLc_s+BL^2c_s^4)$, which contains all the $\theta_s$ dependence in Eq.~(\ref{Chi2Dom}) and recognizing that $B$ and $C$ are small numbers. 
\item It is easiest to measure the CP-odd phase if $\phi_s = \pm \pi/2$. This can be seen in $g(\phi_s,\phi^{\star}_s)$, the maximal variance of this is for $\phi^{\star}_s = \pi/2$ and $\phi_s = -\pi/2$, or vice-versa.
\end{itemize}
We note that these observations do not depend on the details of the beam nor the detector but rather stem from the form of the oscillation probability. 

The best strategy for choosing a baseline $L$ to maximize sensitivity for measuring $|U_{e4}|^2|U_{\mu4}|^2$ and $\Delta m_{14}^2$ is to require that the highest-energy (first) oscillation maximum is within the energy range associate with the experiment.  If so, the signal yield mostly depends on $|U_{e4}|^2|U_{\mu4}|^2$, and the peak of the measured signal distribution mostly depends on $\Delta m_{14}^2$.  On the other hand, the best strategy for measuring the effects of CPV -- the subject of our study -- is to arrange for multiple oscillations to occur within the measured neutrino energy range. These may or may not include the first oscillation maximum.

\begin{figure}[htbp]
\begin{center}
\subfigure[]{\label{prob1}\includegraphics[width=0.6\textwidth]{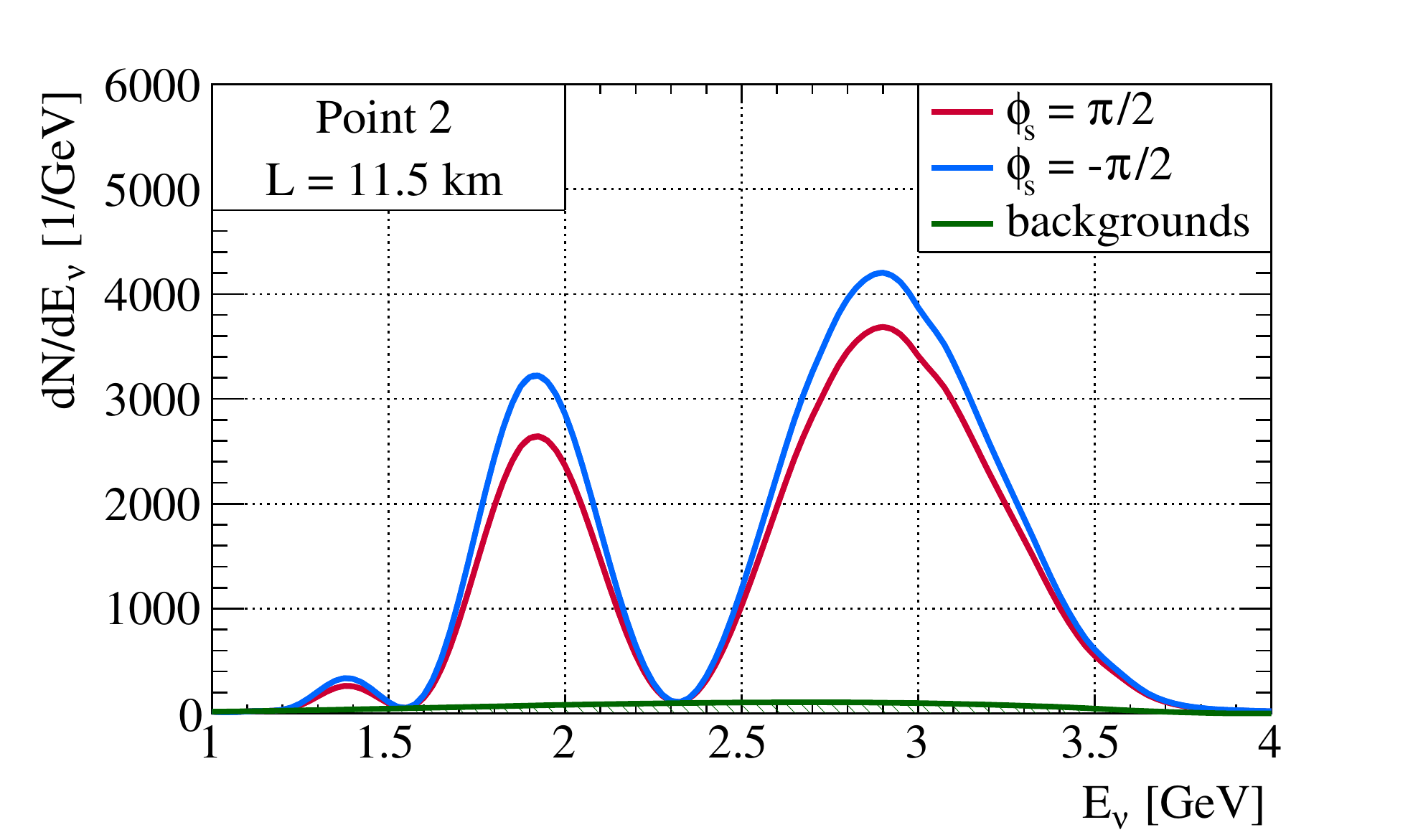}}
\hspace{0.5in}
\subfigure[]{\label{prob2}\includegraphics[width=0.6\textwidth]{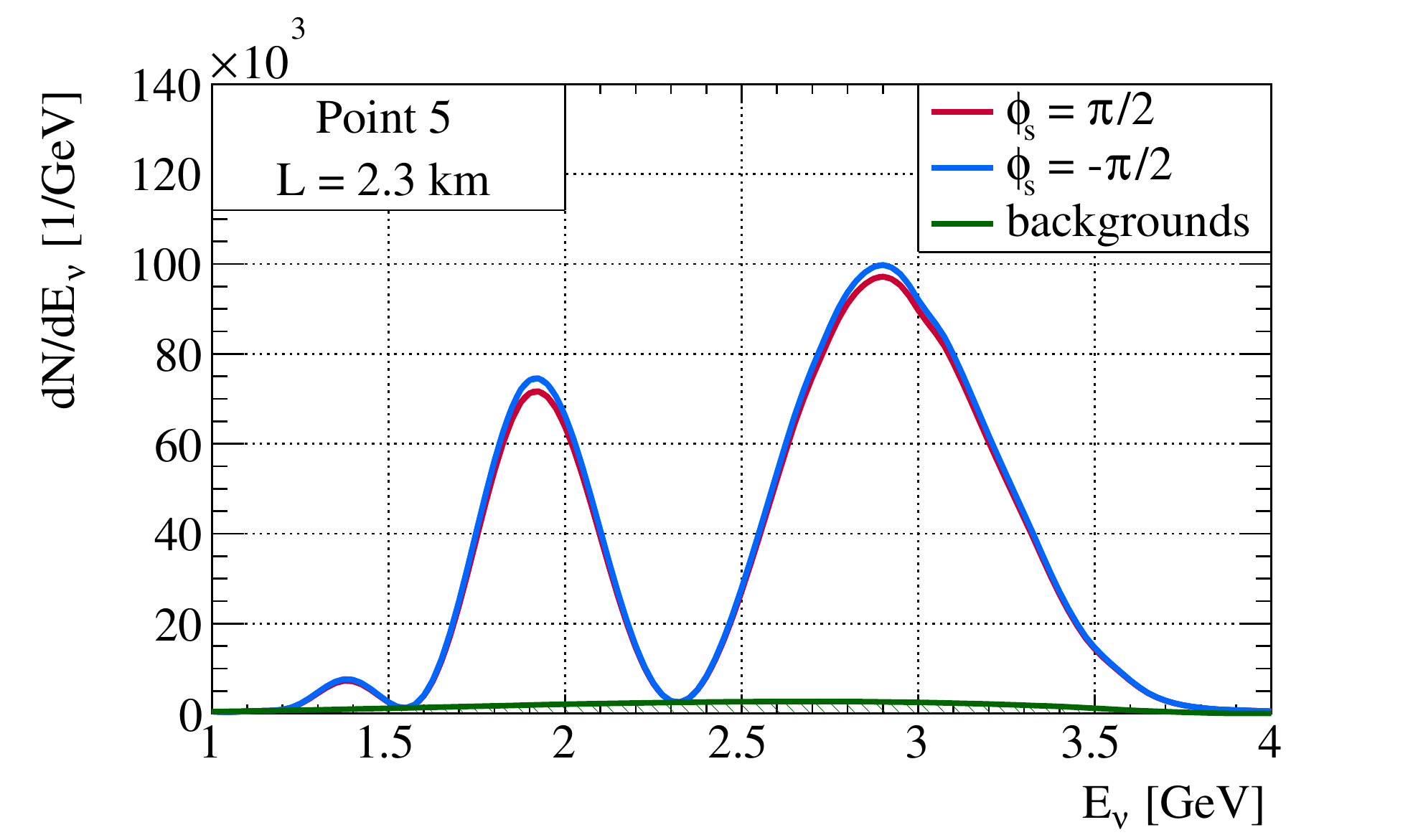}}
\subfigure[]{\label{prob3}\includegraphics[width=0.6\textwidth]{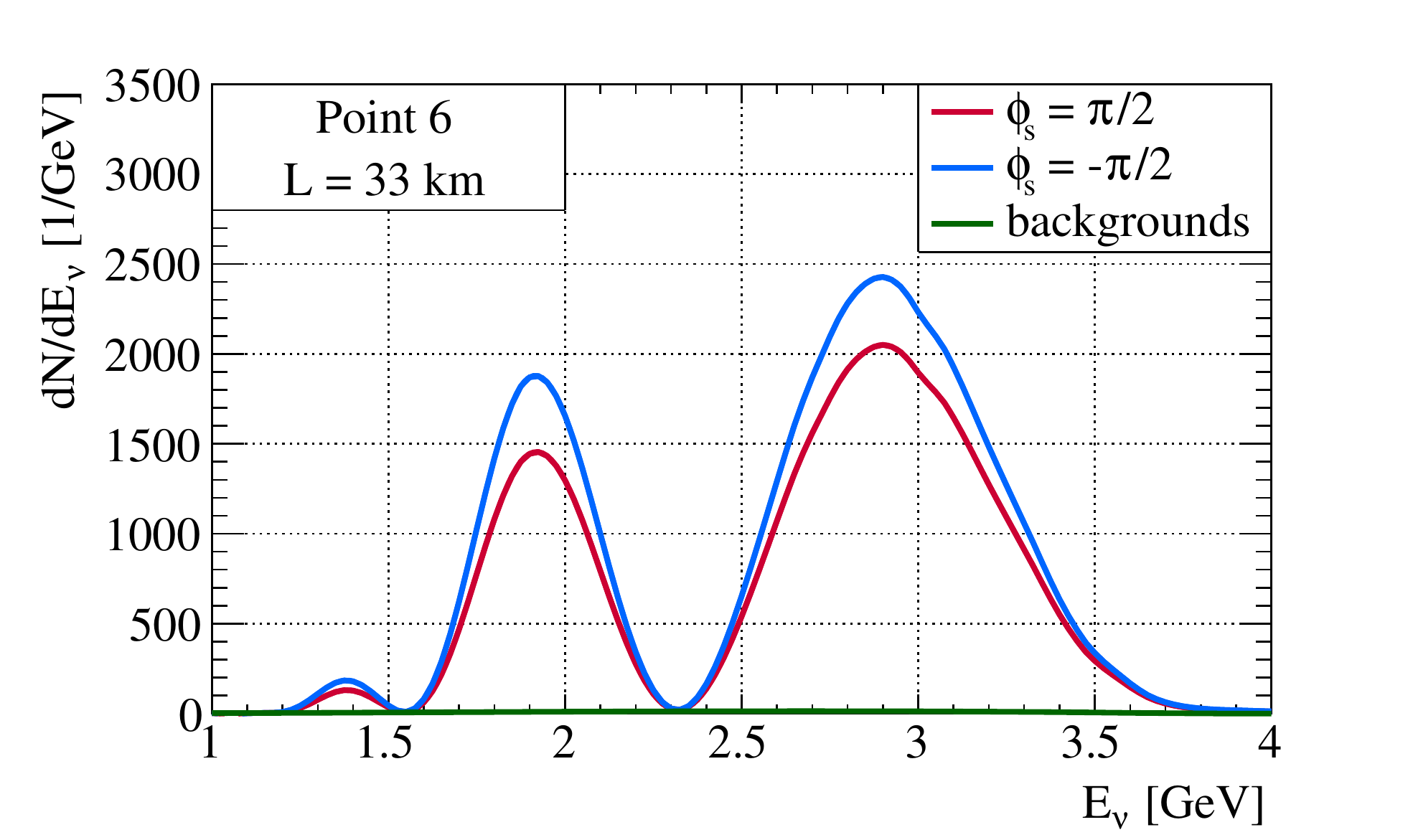}}
\caption{The differential number of expected $\nu_e\rightarrow\nu_\mu$ events  as a function of the neutrino energy at $\nu$STORM+. Shown are the expected yields when $\phi_s=\pm\pi/2$. Backgrounds correspond to misidentified charged-current and neutral-current beam-induced events from $\overline{\nu}_\mu\rightarrow \overline{\nu}_\mu$ and $\nu_e \rightarrow \nu_e$ \cite{Adey:2014rfv}. The three panels correspond to (a) Point 2, at $L=11.5$ km, (b) Point 5, at $L=2.3$ km, and (c) Point 6, at $L=33$ km. See text for detais.}
\label{probs}
\end{center}
\end{figure}

Based on the discussion of the $\chi^2$ function, the best value of $\Delta m_{14}^2L$ can be estimated independent from the physical value of $\Delta m_{14}^2$.  For the $\nu$STORM+ flux shape, signal efficiency, and background rate, we find that choosing the product $\Delta m_{14}^2 L$ equal to $11.5$ eV$^2\cdot$km optimizes the sensitivity to $\phi_s$. This result is obtained by calculating the maximal $\Delta \chi^2$ for a particular set of parameters and varying $L$.  We find, for all Points 1-6 in the parameter space, that the greatest sensitivity for measuring the effects of CPV is for a fixed value of $\Delta m_{14}^2 L$.  If a different beam profile were chosen, then this value would change. We illustrate this fact by concentrating on Points 2, 5, and 6, where $\Delta m_{14}^2 = 1.0$ eV$^2$, 5.0 eV$^2$, and 0.35 eV$^2$, respectively. Fig.~\ref{probs} depicts the ``data'' corresponding to Points 2, 5, and 6 for $L=11.5$ km, 2.3 km, and 33 km, respectively, so $\Delta m_{14}^2 L = 11.5$ eV$^2\cdot$km for all three panels, and $\phi_s=\pm\pi/2$. It is easy to see that, while the number of events and the relative CPV effects are quite different, the ``shapes'' corresponding to the three points are almost identical. As we will show immediately, the sensitivity to $\phi_s$ is almost identical for these three scenarios.

\begin{figure}[htbp]
\begin{center}
\includegraphics[width=0.7\textwidth]{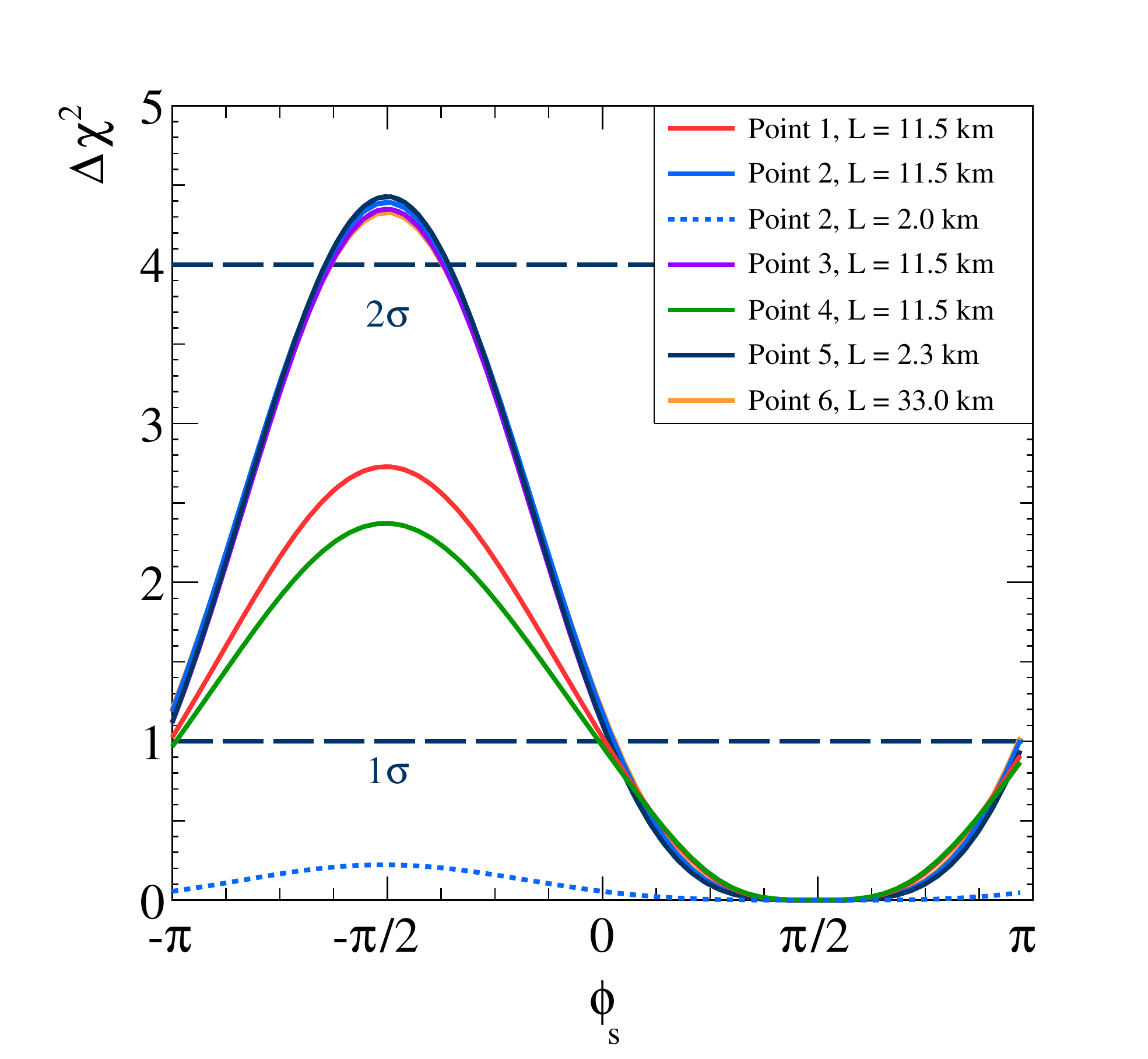}
\caption{The expected values of $\Delta \chi^2$ at $\nu$STORM+ when $\phi_s=\pi/2$ for the six points shown in Fig.~\ref{nustormplot} and Table~\ref{table:nustormprecision}. The dotted line correspond to Point 2, assuming $\nu$STORM+ is performed at the $\nu$STORM baseline.}
\label{phisensitivity}
\end{center}
\end{figure}

Our ``measurements'' of $\phi_s$ are depicted in Fig.~\ref{phisensitivity}, for the six values of $4|U_{e4}|^2|U_{\mu4}|^2$ and $\Delta m_{14}^2$ in Fig.~\ref{nustormplot} and Table~\ref{table:nustormprecision}, for $\Delta m_{14}^2 L = 11.5$ eV$^2\cdot$km and $\phi_s = \pi/2$.  We make use of Eq.~(\ref{Chi2}) and compute $\Delta\chi^2$  after numerically marginalizing over $\Delta m^2_{14}$ and $\theta_s$, for each Point. In the Appendix, we present constant $\chi^2$ contours in the two-dimensional $\phi_s\times \Delta m^2_{14}$ and  $\phi_s\times \theta_s$ planes, for Point 2. As advertised, if $\theta_s$ is not much larger nor smaller than 0.18, i.e., $4|U_{e4}|^2|U_{\mu4}|^4\sim4\times10^{-3}$, the value of $\Delta\chi^2$ changes minimally for different values of $\Delta m_{14}^2$, as long as $\Delta m_{14}^2 L = 11.5$ eV$^2\cdot$km is satisfied.  Qualitatively, for different values of $\phi_s$, one obtains similar results, though the overall sensitivity for measuring $\phi_s$ is reduced. Also depicted in Fig.~\ref{phisensitivity} (dotted line), is the measurement of $\phi_s$ in the case nature agrees with Point 2 but assuming $\nu$STORM+ is performed at the proposed $\nu$STORM baseline. It is apparent that the optimal choice of baseline is very significant. 

$\nu$STORM+ can establish, for all points defined in Fig.~\ref{nustormplot}, that CP-invariance is violated if the CP-odd phase is $\pi/2$ only at the one-sigma level, i.e., it would constrain $\phi_s\in [0,\pi]$ at the one-sigma level and can rule out $\phi_s=-\pi/2$ at around the two-sigma level (Fig.~\ref{phisensitivity}).  As discussed above, the expected uncertainties are larger for different values of $\phi_s$. More power to establish CPV might come from combining other information on $\Delta m^2_{14}$, $\theta_s$, and $\phi_s$. Such information may come from the disappearance channel, combining neutrino running with antineutrino running, or combining searches for $\nu_e\to \nu_{\mu}$ with those for $\nu_{\mu}\to\nu_e$. The latter could be pursued, for example, at different experiments making use of a well-characterized $\nu_{\mu}$ source, including pion decay at rest \cite{Alonso:2010fs}. The $\nu$STORM+ disappearance data from the process $P_{\bar{\mu}\bar{\mu}}$ would be available concurrently with those from $P_{e\mu}$. While they provide no information on $\theta_s$ -- $\nu_{\mu}$ disappearance is mostly dependent on $\theta_{24}$, providing virtually no information on $\theta_{14}$ -- they do provide a different measurement of $\Delta m^2_{14}$, mostly independent from $\phi_s$. Performing a joint appearance and disappearance analysis is beyond the scope of this paper.\footnote{One challenge is that, once appearance and disappearance data are combined, the $\theta_{14}=\theta_{24}=\theta_s$ choice can no longer be made, and one is required to explore the three dimensional $\theta_{14},\theta_{24},\Delta m^2_{14}$ new-physics parameter space. This renders the discussion much more cumbersome.} Nonetheless, we estimate the consequences of combining the two data sets by ``measuring'' $\Delta m^2_{14}$ using disappearance data and applying the result as a prior to the $\chi^2$ analyses described in detail in this section. For all Points, except Point 6, we find that one can exclude $\phi_s=0$ at a level somewhere between two and three sigma ($4\lesssim \Delta \chi^2\lesssim 10$, depending on the Point). An improved measurement of $\theta_s$ might prove at least as fruitful. In this case, however, in order to make use of disappearance data, one needs to combine both $\nu_{\mu}$ and $\nu_e$ disappearance in order to determine both $\theta_{14}$ and $\theta_{24}$. A precise measurement of $\nu_e$ disappearance would require, for example, a short-baseline reactor experiment (see, for example, \cite{Ashenfelter:2013oaa}) or a radioactive-source experiment (see, for example, \cite{Bungau:2012ys}).

\setcounter{equation}{0}
\section{Conclusion}
\label{sec:conclusion}

The discovery of new, light neutrino degrees of freedom would qualitatively impact our understanding of fundamental physics. A new wave of oscillation experiments, aimed at exploring the physics at the new oscillation length(s), will be required in order to explore the new-physics sector. 

New neutrino degrees of freedom allow for more sources of CP-invariance violation (CPV). Here, we explore the requirements for accessing CP-odd mixing parameters in the so-called $3+1$ scenario, where one assumes the existence of one extra, mostly sterile neutrino degree of freedom, significantly heavier than the other three mass eigenstates. CPV is present in the interference term between the solar and atmospheric oscillation lengths, proportional to $\Delta m^2_{12}$ and $\Delta m^2_{13}$ respectively, and the new shorter oscillation length, proportional to $\Delta m^2_{14}$. We concentrate on short-baseline experiments, engineered such that $\Delta m^2_{14}\sim E_{\nu}/L$, and argue that solar effects, due to the fact that $\theta_{13}$ is not too small, can be safely neglected. We also show that, if new neutrino states are indeed discovered in the next round of short-baseline experiments, atmospheric effects are small, rendering the study of CPV most challenging. Our results confirm that, for the on-going and planned short-baseline experiments, it is safe to approximate $\Delta m^2_{13}=\Delta m^2_{12}=0$ when discussing the $3+N$ oscillation hypotheses, for $N\ge 1$. 

As a first step towards understanding how to measure CPV in short-baseline experiments, we concentrate on the $\nu_{e}\to\nu_{\mu}$ appearance channel in a hypothetical, upgraded version of the $\nu$STORM proposal, $\nu$STORM+. Using only appearance data, we establish that the optimal baseline for CPV studies depends strongly on the value of $\Delta m^2_{14}$ and, in turn, the ability of $\nu$STORM+ to observe CPV depends significantly on whether the experiment is performed at the optimal baseline. 

Our results, assuming a set-up one thousand times more powerful than that of $\nu$STORM ($\nu$STORM+), are depicted in Fig.~\ref{phisensitivity}. Even at the optimal baselines, it will be very challenging to see CPV in $3+1$ scenarios if one considers only one appearance channel. Significantly better results are expected if one includes more information. Some is already accessible at $\nu$STORM+, including more information on $\Delta m^2_{14}$ from the $\nu_{\mu}$ disappearance channel. Other possibilities include combining the neutrino and the antineutrino appearance channels by changing the charge of the muons in the storage ring, or combining $\nu_{e}\to\nu_{\mu}$ data with those from a different experiment capable of precision measurements of $\nu_{\mu}\to\nu_{e}$, the T-conjugate channel. 

Even in the simple $3+1$ scenario, CPV effects beyond those studied here can be easily accommodated. The study of the tau-appearance channel ($\nu_{\mu}\to \nu_{\tau}$ or $\nu_{e}\to \nu_{\tau}$) is required for exploring the second new CP-odd ``Dirac'' phase contained in the extended $4\times 4$ mixing matrix. As of right now, if there is a new mass-squared difference of order 1~eV$^2$, very little is known about the $\nu_{\tau}$ content of the fourth neutrino mass eigenstate. Searches for tau-appearance -- let alone precision measurements of tau-appearance -- are extremely challenging and will require new, dedicated experimental efforts that go significantly beyond what is currently being explored by the experimental neutrino community.

\section*{Acknowledgements}

This work is sponsored in part by the DOE grant \#DE-FG02-91ER40684. 

\bibliography{SterileBib}{}

\appendix
\setcounter{equation}{0}
\section{Determining the New Oscillation Parameters: Point 2}

\begin{figure}[htbp]
\begin{center}
\subfigure{\includegraphics[width=0.48\textwidth]{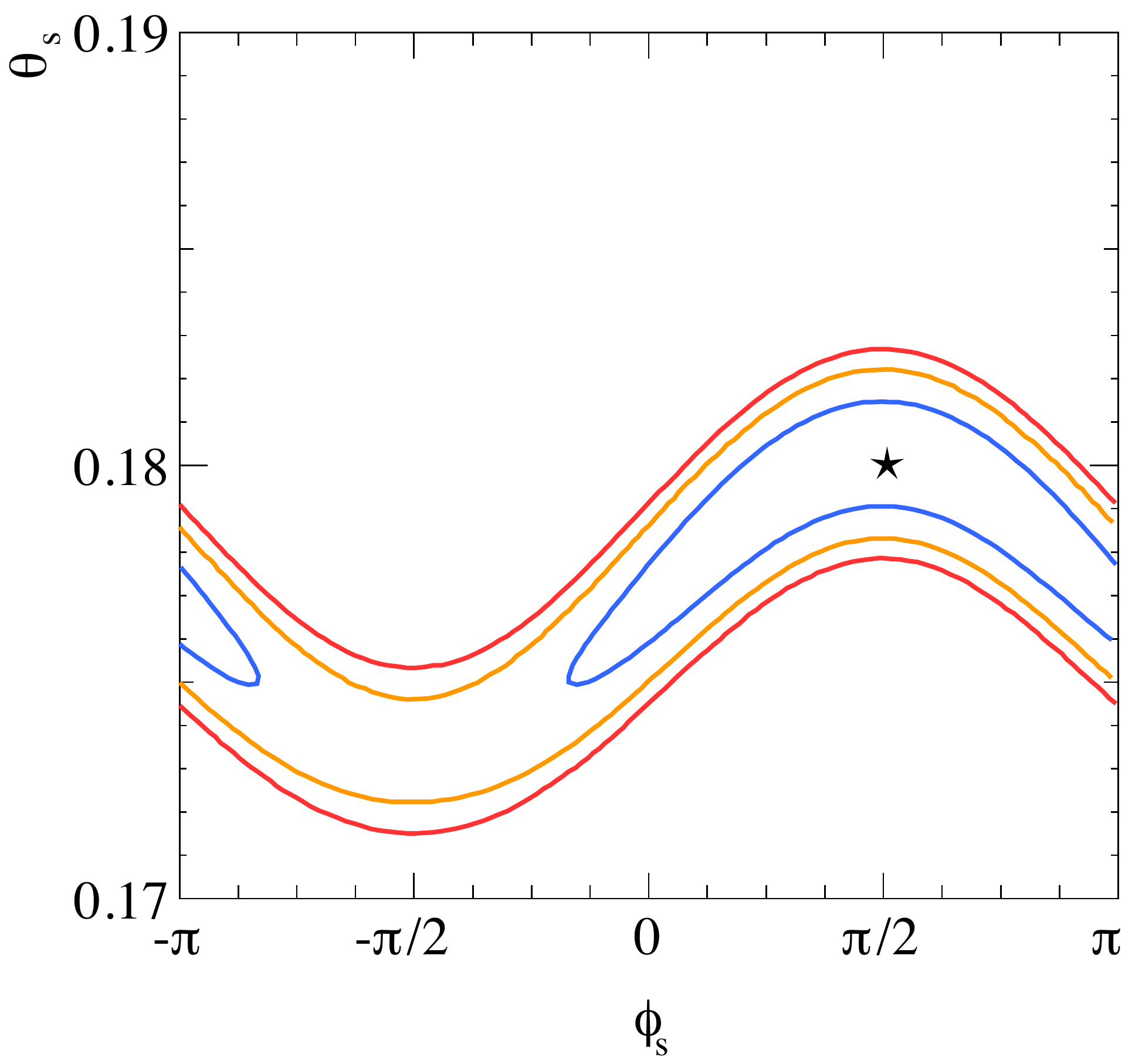}}
\subfigure{\includegraphics[width=0.48\textwidth]{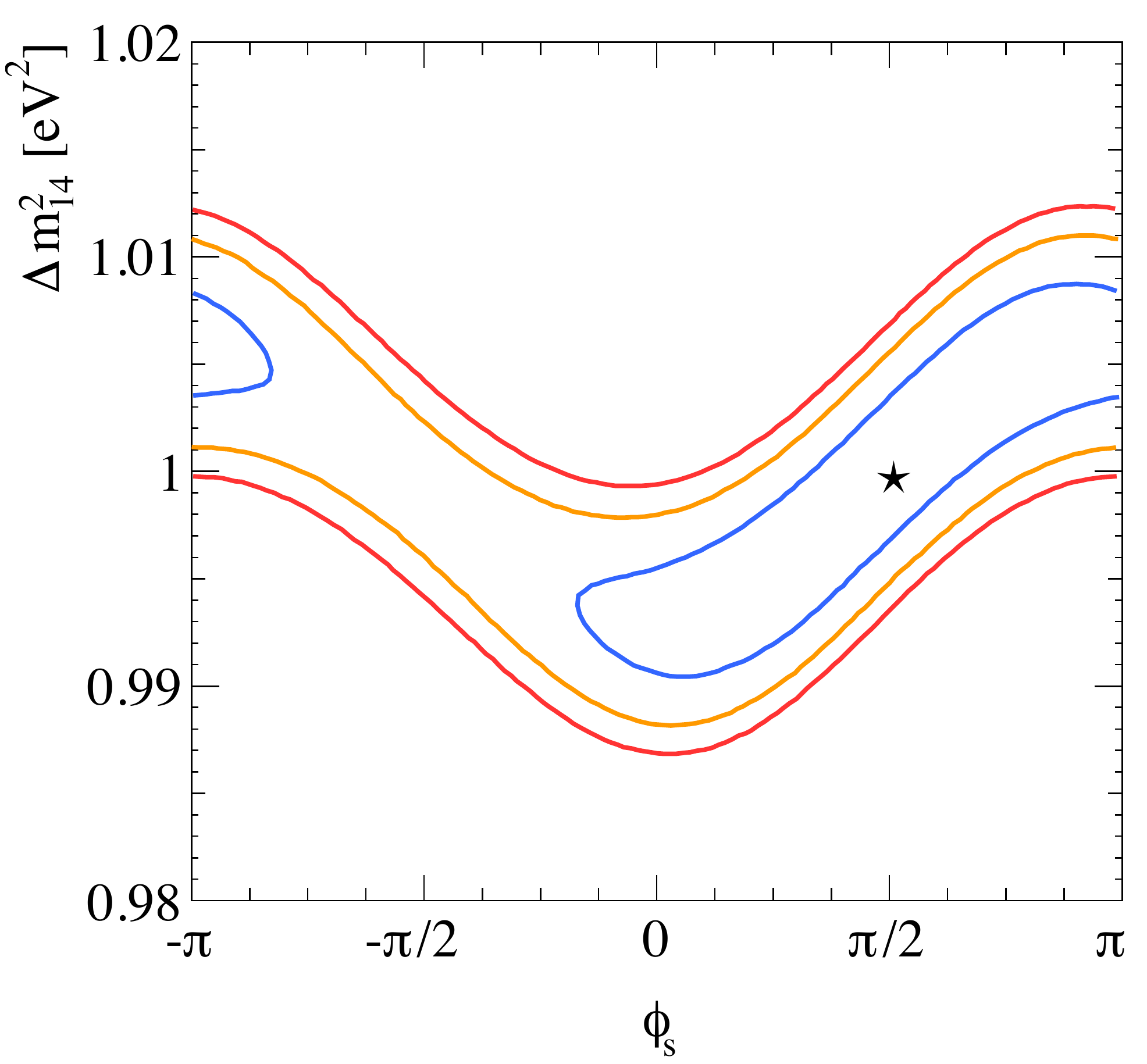}}
\caption{Sensitivity contours corresponding to Point 2 in Fig.~\ref{nustormplot} and Table~\ref{table:nustormprecision}. Blue: 68\%~CL, Yellow: 95\%~CL, Red: 99\%~CL. ``Physical'' values are indicated by a star in each plot. The third parameter is marginalized over in each panel.}
\label{AppendixFig}
\end{center}
\end{figure}

Fig.~\ref{AppendixFig} depicts the experimental sensitivity to the oscillation parameters at the $\nu$STORM$+$ experiment, outlined in Section~\ref{sec:exp}, for physical values (i.e. input values) of the parameters $\Delta m_{14}^2 = 1.0$ eV$^2$, $4|U_{e4}|^2|U_{\mu 4}|^2 = 4\times 10^{-3}$ (or $\theta_s = 0.18$), and $\phi_s = \pi/2$ (Point 2 in Fig.~\ref{nustormplot} and Table~\ref{table:nustormprecision}). In each plot, the third variable is marginalized over when calculating $\Delta \chi^2$ contours. The blue, yellow, and red contours correspond to $68\%$, $95\%$, and $99\%$ CL sensitivity, respectively. Stars indicate the input values of $\phi_s$, $\theta_s$, and $\Delta m_{14}^2$.

\end{document}